\documentclass{iau} 
\usepackage{graphicx}
\usepackage{amsmath}

\title[Asteroseismology of magnetic cycles] 
{Seismic signatures of magnetic activity in solar-type stars observed by \textit{\textbf{Kepler}}}

\author[A. R. G. Santos et al.]   
{A. R. G. Santos$^{1,2,3,4}$, T. L. Campante$^{2,3,4}$, W. J. Chaplin$^{4,5}$, M. S. Cunha$^{2,3}$, M. N. Lund$^{4,5}$, R. Kiefer$^6$, D. Salabert$^{7,8}$, R. A. Garc\'{i}a$^{7,8}$, G. R. Davies$^{4,5}$, Y. Elsworth$^{4,5}$, and R. Howe$^{4,5}$}

\affiliation{
$^1$ Space Science Institute, 4750 Walnut Street, Suite 205, Boulder CO 80301, USA \\[\affilskip]
$^2$ Instituto de Astrof\'{i}sica e Ci\^{e}ncias do Espa\c{c}o, Universidade do Porto, CAUP, Rua das Estrelas, PT-4150-762 Porto, Portugal\\[\affilskip]
$^3$ Departamento de F\'{i}sica e Astronomia, Faculdade de Ci\^{e}ncias, Universidade do Porto, Rua do Campo Alegre 687, PT-4169-007 Porto, Portugal\\[\affilskip]
$^4$ School of Physics and Astronomy, University of Birmingham, Edgbaston, Birmingham B15 2TT, UK\\[\affilskip]
$^5$ Stellar Astrophysics Centre, Department of Physics and Astronomy, Aarhus University, Ny Munkegade 120, DK-8000 Aarhus C, Denmark\\[\affilskip]
$^6$ Kiepenheuer-Institut f\"{u}r Sonnenphysik, Sch\"{o}neckstra\ss e 6, 79104 Freiburg, Germany \\[\affilskip]
$^7$ IRFU, CEA, Universit\'e Paris-Saclay, F-91191 Gif-sur-Yvette, France \\[\affilskip]
$^8$ Universit\'e Paris Diderot, AIM, Sorbonne Paris Cit\'e, CEA, CNRS, F-91191 Gif-sur-Yvette, France
}

\pubyear{2018}
\volume{340}  
\jname{Long-term datasets for the understanding of solar and stellar magnetic cycles}
\editors{A.C. Editor, B.D. Editor \& C.E. Editor, eds.}
\begin{document}

\maketitle

\begin{abstract}
The properties of the acoustic modes are sensitive to magnetic activity. The unprecedented long-term {\it Kepler} photometry, thus, allows stellar magnetic cycles to be studied through asteroseismology. We search for signatures of magnetic cycles in the seismic data of {\it Kepler} solar-type stars. We find evidence for periodic variations in the acoustic properties of about half of the 87 analysed stars. In these proceedings, we highlight the results obtained for two such stars, namely KIC~8006161 and KIC~5184732.

\end{abstract}

\firstsection 
\section{Introduction}

As a result of the magnetic activity, the properties of the solar acoustic modes are observed to vary periodically. In particular, mode frequencies and amplitudes show a temporal anti-correlation, with frequencies increasing with increasing activity, while amplitudes decrease (e.g.,\cite[Woodard \& Noyes 1985]{Woodard}; \cite[Elsworth \etal\ 1990]{Elsworth1990}; \cite[Libbrecht \& Woodard 1990]{Libbrecht1990}; \cite[Chaplin \etal\ 1998]{Chaplin1998}; \cite[Howe \etal\ 2015]{Howe2015}). 
\cite[Garc\'{i}a \etal\ (2010)]{Garcia2010} detected, for the first time, activity-related variations in the acoustic properties of a star other than the Sun: HD~49933 observed by CoRoT. Taking advantage of the long-term {\it Kepler} photometric time-series, temporal frequency shifts, possibly activity-related, have been measured for a number of solar-type stars (\cite[Salabert \etal\ 2016, 2018]{Salabert2016,Salbert2018}; \cite[R\'{e}gulo \etal\ 2016]{Regulo2016}; \cite[Kiefer \etal\ 2017]{Kiefer2017}).

We have searched for temporal variations in the seismic properties of {\it Kepler} solar-type stars. To that end, we developed a Bayesian peak-bagging tool. In these proceedings, we summarize the methodology and highlight the results for two stars in the target sample.

\section{Observational data}

We analysed {\it Kepler} short-cadence data for 87 solar-type stars. 
The pixel data were collected from KASOC ({\it Kepler} Asteroseismic Science Operations Center) and corrected using the KASOC filter (\cite[Handberg \& Lund 2014]{Handberg2014}). The time series were then split in 90-day sub-series and, for each, the power density spectrum is obtained. For the photometric activity proxy, $S_{\rm ph}$ (e.g., \cite[Mathur \etal\ 2014]{Mathur2014}), we use KADACS ({\it Kepler} Asteroseismic Data Analysis and Calibration Software; \cite[Garc\'{i}a \etal\ 2011]{Garcia2011}) long-cadence light curves.

\section{Modelling of the power density spectrum}

Stellar brightness varies on different timescales, due to the contribution from different phenomena, such as magnetic features, granulation, and stellar oscillations. 
We start by describing the background signal as the sum of three components: an exponential decay of active regions; a Harvey-like profile for granulation; and a constant photon shot-noise. 

Having the background model, we proceed with the peak-bagging analysis and perform a global fit of the acoustic modes. For each mode, the power spectrum is modelled as a Lorentzian profile. The final set of free parameters is composed of the mode frequencies $\nu_{nl}$ (where $n$ and $l$ denote the radial order and angular degree), the heights and linewidths of the radial modes, the rotational splitting, and the stellar inclination angle.

To finally obtain the mode parameters, we adopt a Bayesian approach. One of the advantages of a Bayesian approach is the possibility of using prior knowledge to constrain the parameters. Thus, in this work, the prior probability functions (namely for mode frequencies, rotational splitting, and inclination) are based on previous results from the analysis of the full, multi-year time-series (\cite[Davies \etal\ 2016; Lund \etal\ 2017]{Davies2016,Lund2017}). 

The optimization method makes use of the algorithm \texttt{emcee} (\cite{Foreman-Mackey2013}), based on the Affine Invariant Markov Chain Monte Carlo Ensemble sampler (\cite{Goodman2010}). From this analysis, we obtain the posterior distribution for each parameter and the corresponding parameter estimates.

\section{Results and Discussion}

Having the mode parameters, we then compute the mean temporal frequency shifts and mode heights. The individual frequency shifts, $\delta\nu_{nl}$, are computed with respect to the reference frequencies (weighted averages of the mode frequencies, $\nu_{nl}$).
The final mean frequency shifts, $\delta\nu$, and uncertainties, $\sigma$, are obtained as
\begin{equation}
\delta\nu(t)=\dfrac{\Sigma_{nl}\delta\nu_{nl}(t)/\sigma^2_{nl}(t)}{\Sigma_{nl}1/\sigma^2_{nl}(t)},
\end{equation}\begin{equation}
\sigma(t)=\left(\Sigma_{nl}1/\sigma_{nl}^2(t)\right)^{-1/2}.
\end{equation}

\vspace{0.2cm}
For the mode heights, we follow the same approach, but using logarithmic values.

\begin{figure}[h]
\includegraphics[width=\hsize]{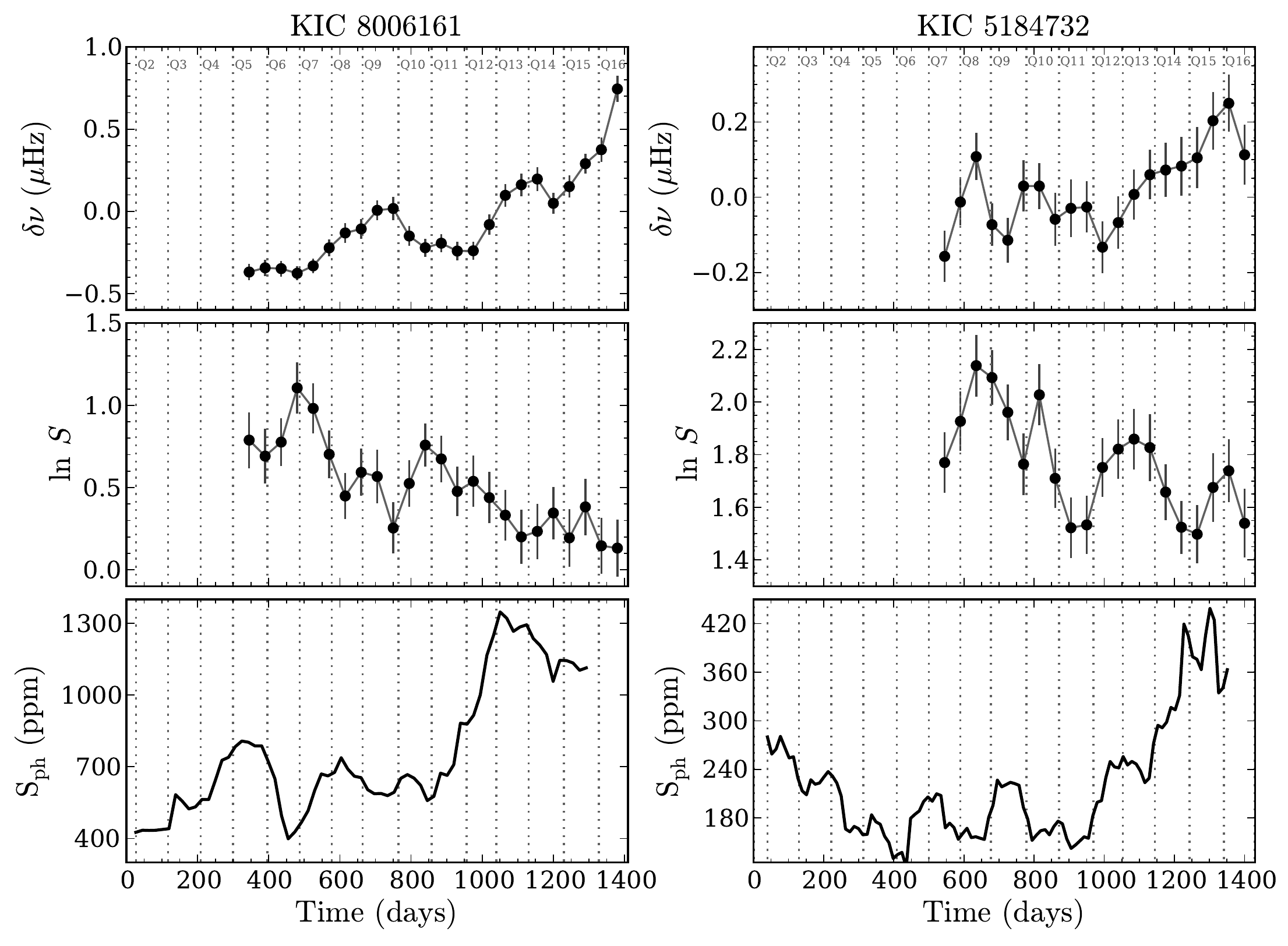}
\caption{Results for KIC~8006161 ({\it left}) and KIC~5184732 ({\it right}). {\it Top and middle:} Frequency shifts and logarithmic mode heights. {\it Bottom:} Photometric activity proxy. Vertical dotted lines mark the {\it Kepler} quarters.}\label{fig:fig01}
\end{figure}

We perform the analysis, summarized above, for all 87 solar-type stars. In these proceedings, in Fig. \ref{fig:fig01}, we present the results for the two stars KIC~8006161 and KIC~5184732. The seismic indicators (frequency shifts, $\delta\nu$, and logarithmic heights, $\ln S$) are shown in the top two rows. For comparison, the bottom panels show the photometric activity proxy, $S_\text{ph}$, which is a measure of the stellar brightness variations due to the presence of spots on the stellar surface. Similarly to what is observed in the Sun, for both stars, the frequency shifts and mode heights vary in anti-phase, with the frequency shifts increasing with increasing $S_\text{ph}$. Therefore, our results are consistent with the rising phase of an activity cycle in both stars. $S_\text{ph}$ further suggests that, over the time span of the {\it Kepler} observations, a given cycle ended and new cycle began in KIC~5184732. A detailed study of the enhanced activity and strong surface differential rotation of KIC~8006161 is presented in \cite{Karoff2018}. 

Over $50\%$ of the stars in the target sample show evidence for quasi-periodic variations in the frequency shifts. For part of those, the frequency shifts are accompanied by variations in other stellar properties, such as mode heights, granulation timescale, and/or photometric activity proxy. Further details will be provided in \cite[Santos \etal\ (2018)]{Santos2018}.

\acknowledgments

This work was supported by Funda\c{c}\~{a}o para a Ci\^{e}ncia e a Tecnologia (FCT) through national funds (UID/FIS/04434/2013) and by FEDER through COMPETE2020 (POCI-01-0145-FEDER-007672). ARGS acknowledges the support from the IAU travel grant, from NASA grant NNX17AF27G, from the fellowship SFRH/BD/88032/2012 funded by FCT (Portugal) and POPH/FSE (EC), and from University of Birmingham. TLC acknowledges support from grant CIAAUP-12/2018-BPD. TLC, WJC, GRD, EY, and RH acknowledge the support from the UK Science and Technology Facilities Council (STFC). MSC acknowledges the support from FCT through the Investigador FCT Contract No. IF/00894/2012 and the fellowship SFRH/BD/88032/2012 and by FEDER through COMPETE2020 (POCI-01-0145- FEDER-007672). MNL acknowledges the support of The Danish Council for Independent Research | Natural Science (Grant DFF-4181-00415). Funding for the Stellar Astrophysics Centre (SAC) is provided by The Danish National Research Foundation (Grant agreement no.: DNRF106). RK acknowledges that the research leading to these results received funding from the European Research Council under the European Unions Seventh Framework Program (FP/2007-2013)/ERC Grant Agreement no. 307117. DS and RAG acknowledge the support from the CNES GOLF grant. The research leading to these results has received funding from EC, under FP7, through the grant agreement FP7-SPACE-2012-312844 (SPACEINN) and PIRSES-GA-2010-269194 (ASK). The peak-bagging was performed using the University of Birmingham's BlueBEAR HPC service (http://www.birmingham.ac.uk/bear).

\end{document}